\documentclass[iop]{emulateapj}
\accepted{2015 May 13}
\slugcomment{}
\shorttitle{230 GHz VLBI observations of M87 in March 2012}
\shortauthors{K. Akiyama et al.}
\usepackage[
dvipdfm,
bookmarks=true,
bookmarksnumbered=true,
bookmarkstype=toc,
colorlinks=true,
]{hyperref}
\usepackage{color}
\newcommand{\utokyo}{1}
\newcommand{\naoj}{2}
\newcommand{\haystack}{3}
\newcommand{\mpifr}{4}
\newcommand{\cfa}{5}
\newcommand{\perimeter}{6}
\newcommand{\uwaterloo}{7}
\newcommand{\ucberkely}{8}
\newcommand{\mpe}{9}
\newcommand{\irainaf}{10}
\newcommand{\kasi}{11}
\newcommand{\sokendai}{12}
\newcommand{\asiaa}{13}
\newcommand{\radboud}{14}
\newcommand{\asiaahi}{15}
\newcommand{\aro}{16}
\newcommand{\jcmt}{17}
\newcommand{\unamm}{18}
\newcommand{\uconcepcion}{19}
\newcommand{\inaoe}{20}	
\newcommand{\leiden}{21}
\begin{document}
%------------------------------------------------------------------------------
% Title
%------------------------------------------------------------------------------
\title{230 GHz VLBI observations of M87: event-horizon-scale structure at the enhanced very-high-energy $\rm \gamma$-ray state in 2012}

%------------------------------------------------------------------------------
% Authors
%------------------------------------------------------------------------------
\author{Kazunori Akiyama\altaffilmark{\utokyo,\naoj,*},
Ru-Sen Lu\altaffilmark{\haystack,\mpifr},
Vincent L. Fish\altaffilmark{\haystack},
Sheperd S. Doeleman\altaffilmark{\haystack,\cfa},
Avery E. Broderick\altaffilmark{\perimeter,\uwaterloo},
Jason Dexter\altaffilmark{\ucberkely},
Kazuhiro Hada\altaffilmark{\naoj,\irainaf},
Motoki Kino\altaffilmark{\kasi},
Hiroshi Nagai\altaffilmark{\naoj},
Mareki Honma\altaffilmark{\naoj,\sokendai},
Michael D. Johnson\altaffilmark{\cfa},
%A
Juan C. Algaba\altaffilmark{\kasi,\asiaa},
Keiichi Asada\altaffilmark{\asiaa},
%B
Christiaan Brinkerink\altaffilmark{\radboud},
Ray Blundell\altaffilmark{\cfa},
Geoffrey C. Bower\altaffilmark{\asiaahi},
%C
Roger Cappallo\altaffilmark{\haystack},
Geoffrey B. Crew\altaffilmark{\haystack},
%D
Matt Dexter\altaffilmark{\ucberkely},
Sergio A. Dzib\altaffilmark{\mpifr, \unamm}
%E
%F
Robert Freund\altaffilmark{\aro},
Per Friberg\altaffilmark{\jcmt},
%G
Mark Gurwell\altaffilmark{\cfa},
%H
Paul T.P. Ho\altaffilmark{\asiaa},
%I
Makoto Inoue\altaffilmark{\asiaa},
%J
%K
Thomas P. Krichbaum\altaffilmark{\mpifr},
%L
Laurent Loinard\altaffilmark{\unamm},
%M
David MacMahon\altaffilmark{\ucberkely},
Daniel P. Marrone\altaffilmark{\aro},
James M. Moran\altaffilmark{\cfa},
%N
Masanori Nakamura\altaffilmark{\asiaa},
Neil M. Nagar\altaffilmark{\uconcepcion},
%O
Gisela Ortiz-Leon\altaffilmark{\unamm},
%P
Richard Plambeck\altaffilmark{\ucberkely},
Nicolas Pradel\altaffilmark{\asiaa},
Rurik A. Primiani\altaffilmark{\cfa},
%Q
%R
Alan E. E. Rogers\altaffilmark{\haystack},
Alan L. Roy\altaffilmark{\mpifr},
%S
Jason SooHoo\altaffilmark{\haystack},
%T
Jonathan-Le\'on Tavares\altaffilmark{\inaoe},
Remo P. J. Tilanus\altaffilmark{\radboud,\leiden},
Michael Titus\altaffilmark{\haystack},
%U
%V
%W
Jan Wagner\altaffilmark{\mpifr,\kasi},
Jonathan Weintroub\altaffilmark{\cfa},
%X
%
Paul Yamaguchi\altaffilmark{\cfa},
Ken H. Young\altaffilmark{\cfa},
%Z
Anton Zensus\altaffilmark{\mpifr}, \and
Lucy M. Ziurys\altaffilmark{\aro}}
%
%
%------------------------------------------------------------------------------
% Affiliations
%------------------------------------------------------------------------------
% Affil: U.Tokyo
\affil{\altaffilmark{\utokyo}Department of Astronomy, Graduate School of Science, The University of Tokyo, 7-3-1 Hongo, Bunkyo-ku, Tokyo 113-0033, Japan}
%
% Affil: NAOJ, Mizusawa VLBI observatory
\affil{\altaffilmark{\naoj}National Astronomical Observatory of Japan, 2-21-1 Osawa, Mitaka, Tokyo 181-8588}
%
% Affil: MIT Haystack observatory
\affil{\altaffilmark{\haystack}Massachusetts Institute of Technology, Haystack Observatory, Route 40, Westford, MA 01886, USA}
%
% Affil: MPIfR
\affil{\altaffilmark{\mpifr}Max-Planck-Institut f\"{u}r Radioastronomie, Auf dem H\"{u}gel 69, D-53121 Bonn, Germany}
%
% Affil: CfA
\affil{\altaffilmark{\cfa}Harvard Smithsonian Center for Astrophysics, 60 Garden Street, Cambridge, MA 02138, USA}
%
% Affil: Perimeter
\affil{\altaffilmark{\perimeter}Perimeter Institute for Theoretical Physics, 31 Caroline Street, North Waterloo, Ontario N2L 2Y5, Canada}
%
% Affil: U.Waterloo
\affil{\altaffilmark{\uwaterloo}Department of Physics and Astronomy, University of Waterloo, 200 University Avenue West, Waterloo, Ontario N2l 3G1, Canada}
%
% Affil: Dept. Astron, UC Berkeley 
\affil{\altaffilmark{\ucberkely}Department of Astronomy, Hearst Field Annex, University of California, Berkeley, CA 94720-3411, USA}
%
% Affil: MPE
\affil{\altaffilmark{\mpe}Max Planck Institute for Extraterrestrial Physics, Giessenbachstr. 1, 85748 Garching, Germany}
%
% Affil: IRA
\affil{\altaffilmark{\irainaf}INAF Istituto di Radioastronomia, via Gobetti 101, 40129, Bologna, Italy}
%
% Affil: KASI
\affil{\altaffilmark{\kasi}Korea Astronomy and Space Science Institute (KASI), 776 Daedeokdae-ro, Yuseong-gu,Daejeon 305-348, Republic of Korea}
%
% Affil: SOKENDAI
\affil{\altaffilmark{\sokendai}Graduate University for Advanced Studies, Mitaka, 2-21-1 Osawa, Mitaka, Tokyo 181-8588}
%
% Affil: ASIAA
\affil{\altaffilmark{\asiaa}Institute of Astronomy and Astrophysics, Academia Sinica, P.O. Box 23-141, Taipei 10617, Taiwan}
%
% Affil: Radboud
\affil{\altaffilmark{\radboud}Department of Astrophysics/IMAPP, Radboud University Nijmegen, P.O. Box 9010, 6500 GL Nijmegen, The Netherlands}
%
% Affil: ASIAA/Hawaii
\affil{\altaffilmark{\asiaahi}Academia Sinica Institute of Astronomy and Astrophysics, 645 N. A\'{o}hoku Place, Hilo, HI 96720, USA}
%
% Affil: ARO/SMT
\affil{\altaffilmark{\aro}Arizona Radio Observatory, Steward Observatory, University of Arizona, 933 North Cherry Avenue, Tucson, AZ 85721-0065, USA}
%
% Affil: JCMT
\affil{\altaffilmark{\jcmt}James Clerk Maxwell Telescope, Joint Astronomy Centre, 660 North A\"{o}hoku Place, University Park, Hilo, HI 96720, USA}
%
% Affil: UNAMM
\affil{\altaffilmark{\unamm}Centro de Radiostronom\'{i}a y Astrof\'{i}sica, Universidad Nacional Aut\'{o}noma de M\'{e}xico, Morelia 58089, Mexico}
%
% Affil: U. Concepcion
\affil{\altaffilmark{\uconcepcion}Astronomy Department, Universidad de Concepci\"{o}n, Concepci\"{o}n, Chile}
%
% Affil: INAOE
\affil{\altaffilmark{\inaoe}Instituto Nacional de Astrof\'{i}sica \'{O}ptica y Electr\'{o}nica (INAOE), Apartado Postal 51 y 216, 72000 Puebla, Mexico}
%
% Affil: Leiden
\affil{\altaffilmark{\leiden}Leiden Observatory, Leiden University, PO Box 9513, 2300 RA Leiden, The Netherlands}
%
% JSPS research Fellow
\altaffiltext{*}{\url{kazunori.akiyama@nao.ac.jp}; Research Fellow of the Japan Society for the Promotion of Science}
%
%------------------------------------------------------------------------------
% Abstract
%------------------------------------------------------------------------------
\begin{abstract}
We report on 230 GHz (1.3 mm) VLBI observations of M87 with the Event Horizon Telescope using antennas on Mauna Kea in Hawaii, Mt. Graham in Arizona and Cedar Flat in California. For the first time, we have acquired 230 GHz VLBI interferometric phase information on M87 through measurement of closure phase on the triangle of long baselines. Most of the measured closure phases are consistent with 0$^{\circ}$ as expected by physically-motivated models for 230 GHz structure such as jet models and accretion disk models.
%Our results favor the approaching-jet-dominated models rather than the counter-jet-dominated models, and disfavor the disk-dominated models with mildly low inclination angles inferred for the M87 jets. 
%Our results favor models that generate the observed emission in the approaching jet rather than the counter jet, and disfavor the disk-dominated emission models with mildly low inclination angles inferred for the M87 jets.
%These results do not rule out the existence of the shadow feature in M87, suggesting that future VLBI observations withhigher sensitivity and many more baselines will be essential in determining the structure of emission near the event horizon of M87.
The brightness temperature of the event-horizon-scale structure is $\sim 1 \times 10^{10}$ K derived from the compact flux density of $\sim 1$ Jy and the angular size of $\sim 40 $ $\rm \mu$as $\sim$ 5.5 $R_{{\rm s}}$, which is broadly consistent with the peak brightness of the radio cores at 1-86 GHz located within $\sim 10^2$ $R_{{\rm s}}$.
%For the accretion disk scenario, it favors the hot accretion models with the electron temperature in inner $\sim 10^2$ $R_{{\rm s}}$ higher than the classical advection-dominated accretion flow (ADAF), such as advection-dominated inflow-outflow.
%Combined with results of low frequency observations, it suggests that the relativistic jet is not significantly accelerated in the inner $\sim 10^2$ $R_{{\rm s}}$ from the jet base.
Our observations occurred in the middle of an enhancement in very-high-energy (VHE) $\rm \gamma$-ray flux, presumably originating in the vicinity of the central black hole. Our measurements, combined with results of multi-wavelength observations, favor a scenario in which the VHE region has an extended size of $\sim$20-60 $R_{{\rm s}}$.
\end{abstract}
%
%------------------------------------------------------------------------------
% Keywords
%------------------------------------------------------------------------------
\keywords{
galaxies: active
---galaxies: individual (M87)
---galaxies: jets
---radio continuum: galaxies
---techniques: high angular resolution
---techniques: interferometric}
%
%------------------------------------------------------------------------------
% Main sections
%------------------------------------------------------------------------------
\section{Introduction \label{sec:introduction}}
Relativistic jets pose many intriguing questions in astrophysics related to their formation process and the production mechanism of high energy particles and photons. The relativistic jet in the radio galaxy M87 is an excellent laboratory for investigating these issues; because of its proximity ($D=$ 16.7$\pm$0.6 Mpc; \citealt{Blakeslee2009}) and the large estimated mass of its central black hole ($M_{{\rm BH}}\sim3-6\times10^{9} M_{\odot}$; \citealt{Macchetto1997,Gebhardt2011,Walsh2013}), the black hole in M87 subtends the second largest angular size of any known black hole (after Sgr A*).

Millimeter/submillimeter-wavelength VLBI is ideally suited to observing M87 on these scales, since the event-horizon-scale structure around the black hole is expected to become optically thin at $\nu \gtrsim$ 230 GHz ($\lambda \lesssim 1.3$ mm), based on the frequency-dependent position of the radio core \citep{Hada2011} and the existence of the sub-millimeter bump in its radio spectrum indicating the opacity transition at $\sim$ 230 GHz \citep{Doi2013}. 

The origin of 230 GHz emission is still an unsettled question. The 230 GHz emission could be dominated by synchrotron emission from either the jet \citep{Zakamska2008,Gracia2009,Broderick2009,Dexter2012} or the accretion disk \citep[][]{Reynolds1996,Di Matteo2003,Nagakura2010,Takahashi2011,Dexter2012} in the regime of radiatively inefficient accretion flow \citep[e.g.][]{Yuan2014} with low mass accretion rate of $<9.2\times10^{-4}$ M$_\odot$ y$^{-1}$ \citep{Kuo2014}. The discovery of the position shift of the radio core along the jet direction at different frequencies \citep{Hada2011} provides strong evidence that the jet emission dominates the emission from the radio core at frequencies at least lower than 43 GHz (=7 mm). However, it is less clear for 230 GHz emission, since the extrapolated location of the 230 GHz radio core coincides with the jet base and/or central black hole within its uncertainty, and thus emission from the accretion disk could dominate.

VLBI observations at such high frequencies ($\lambda \lesssim 1.3$ mm, $\nu \gtrsim$ 230 GHz) have been technically challenging due to the limited sensitivity of the instruments, fast atmospheric phase fluctuations and the small number of stations available. Recent technical developments (e.g. phased-array processors, digital backends and recording systems with broad bandwidths) and the addition of new (sub)millimeter telescopes have led to a breakthrough to (sub)millimeter VLBI observations. In particular, significant progress on 230 GHz VLBI observations has been achieved in the last few years with the Event Horizon Telescope \citep[EHT;][]{Doeleman2008,Doeleman2009,Doeleman2012,Fish2011,Fish2013,Lu2012,Lu2013,Lu2014}.

Previous 230 GHz VLBI observations \citep[][hereafter D12]{Doeleman2012} with the EHT established the existence of compact structures on scales of few Schwarzschild radii ($R_{{\rm s}}$), broadly consistent with a paraboloidal or possibly conical collimation profile of the jet width in the innermost region within $\sim$ 100 $R_{\rm s}$ of the central black hole  \citep{Asada2012,Nakamura2013,Hada2013}. These are naturally explained by recent theoretical magnetohydrodynamic (MHD) schemes \citep[e.g.][]{McKinney2006,Komissarov2007}.

VLBI observations at 230 GHz can address at least two issues concerning the fundamental nature of M87. The first is the event-horizon-scale structure of the jet launching region, which is crucial for understanding the formation process of the relativistic jets and also for testing the presence of signatures of strong-field gravitational lensing.  Geometric models including a shadow feature at the last photon orbit, illuminated by a counter jet and/or accretion disk in the close vicinity of the black hole, can be fit to current 230 GHz observations. These models produce a relatively dim central region encircled by a brighter annulus \citep[e.g.][]{Broderick2009,Dexter2012}, which can be properly imaged as the number of (sub)millimeter VLBI sites increases \citep{Lu2014,Honma2014,Inoue2014}.

The second issue is the production mechanism of very-high-energy (VHE; $>$ $\sim$100 GeV) $\rm \gamma$-ray photons in the vicinity of the black hole and/or the jet base. M87 is one of only four known AGNs with weak or moderate beaming compared to other VHE AGNs, which mostly consist of BL Lac objects. M87 has undergone three large VHE flares \citep[see][for an overview]{Abramowski2012} and a weak VHE enhancement recently in March 2012 \citep[][]{Beilicke2012}. In the past three flares, the compact sizes of VHE emission region ($<5\times10^{15}\delta$ cm corresponding to a few $R_{{\rm s}}$, where $\delta$ is the Doppler factor of the emission region) are required by rapid variability timescales of $\sim$1 d based on causality arguments. 
The VHE flares in 2008 and 2010 were followed by delayed strong and weak 43 GHz flux density enhancements, respectively, in the radio core at 43 GHz \citep{Acciari2009,Hada2012}, indicating that these flares originate inside the radio core at 43 GHz only a few tens of $R_{{\rm s}}$ downstream from the black hole and/or jet base \citep{Hada2011}.

On the other hand, a weak VHE enhancement in March 2012 (hereafter the 2012 event) has different properties compared to previous VHE flares. Its long duration ($\sim$2 months) and weak flux ($\sim$10 times weaker than the past three flares) may point to an origin in a different type of VHE activity. Multi-wavelength observations on milliarcsecond scales revealed strong enhancement in the radio core at both 22 and 43 GHz after the 2012 event, suggesting an origin close to the black hole and/or jet base, similar to the 2008 VHE flare \citep[][hereafter H14]{Hada2014}. In summary, three of four previous VHE events are thought to originate in the vicinity of the black hole. 230 GHz VLBI is the ideal tool to constrain the location and structure of the VHE emission region.

We report on new 230 GHz VLBI observations of M87 with the EHT during the 2012 event using a four-telescope array, providing the interferometric visibility information on baselines shorter than $\sim4\,{\rm G \lambda}$. These observations provide the first measurements of closure phase, imposing new constraints on accretion/jet models for M87, and the first constraints on the innermost structure of the relativistic jet on scales of a few $R_{{\rm s}}$ during VHE variability. In this paper, we adopt a black hole mass of $6.2 \times 10^{9}$ M$_{\odot}$\footnote{This black hole mass is recalculated for a distance of 16.7 Mpc.} following \citet{Gebhardt2011} and a distance of 16.7 Mpc following \citet{Blakeslee2009} along with D12, resulting in $1R_{\rm s}=1.9\times10^{15}$ cm $=5.9\times 10^{-4}$ pc $=7.3$ ${\rm \mu}$as.

\section{Observations \label{sec:observations}}
\begin{table*}[t]
\caption{Observatories in the 2012 Observations\label{tab:Telescopes Summary}}
\centering{}%
\begin{tabular}{lllll}
\hline 
\hline Site & Observatory & Char. & Note\\
\hline 
Hawaii & SMA & P & Phased sum of seven 6m dishes\\
Arizona & ARO/SMT & S & Single 10 m dish\\
California & CARMA (phased) & F & Phased sum of three 10.4 m and four 6.1 m dishes\\
California & CARMA (single) & D & Single 10.4 m dish\\
\hline 
\end{tabular}
\end{table*}

%\begin{figure}
%centering{}\includegraphics[width=1\columnwidth]{figure1.eps}
%\caption{
%The correlation co-efficient of M87 derived with and without phase referencing.
%\label{fig:corr-coeff}}
%\end{figure}

M87 and several calibrator sources were observed with four stations at three sites in 2012 on the nights of March 15, 20 and 21 (days 75, 80 and 81), as summarized in Table \ref{tab:Telescopes Summary}: a phased array of the Submillimeter Array (SMA; \citealt{Ho2004}; henceforth, P) antennas and the James Clerk Maxwell Telescope (JCMT; \citealt{Newport1986}) on Mauna Kea in Hawaii, the Arizona Radio Observatory's Submillimeter Telescope (ARO/SMT; \citealt{Martin1986}; S) on Mt.\ Graham in Arizona, and both a single antenna and a phased array of eight antennas of the Combined Array for Research in Millimeter-Wave Astronomy (CARMA; \citealt{Mundy2000}; D and F, respectively) on Cedar Flat in California.

Observations were performed at two bands centered at 229.089 and 229.601 GHz (low and high band) with 480 MHz bandwidths with the exception of the single CARMA antenna, which observed only the low band. All telescopes observed left-hand circular polarization (LHCP). 
The SMT and phased CARMA, along with the JCMT on Mauna Kea, also observed right-hand circular polarization (RHCP).
Hydrogen masers were used as timing and frequency references at all sites. Reconfigurable Open Architecture Computing Hardware (ROACH)\footnote{\url{https://casper.berkeley.edu/wiki/ROACH}} digital backends (RDBE) designed at MIT Haystack Observatory and National Radio Astronomy Observatory (NRAO) were used for all single-antenna stations. Data were recorded onto modules of hard drives using the Mark 5C for RDBE systems. The SMA and CARMA sites were equipped with 1 GHz bandwidth adaptive beamformers, built using an older generation of Collaboration for Astronomy Signal Processing and Electronics Research (CASPER)\footnote{\url{https://casper.berkeley.edu}} technology.  The beamformers compensate group delay and phase at each antenna in the array in real time, thereby recording a single data stream representing the coherent phased array sum of all antennas. The real time corrections are derived from simultaneous cross-correlations, and the data are formatted for Mark5B+ recorders at 4 Gb/s rate. Data were correlated with the Haystack Mark 4 VLBI correlator.

%Data were recorded onto modules of hard drives using the Mark 5B+ for DBE1 systems and Mark 5C for RDBE systems.

Hardware and disk failures occurred during observations on the first two days, with the result that many data products are missing or have low signal-to-noise ratio (S/N). The LHCP data of the first two days and RHCP data can not be calibrated by the technique of amplitude self-calibration described below. In this paper, we focus on the results of LHCP data of M87 in day 81; other data will be presented elsewhere.

\section{Data reduction}
Correlated data were analyzed using the Haystack Observatory Post-processing System\footnote{\url{http://www.haystack.mit.edu/tech/vlbi/hops.html}} (HOPS).
Initial coherent baseline fringe fitting was done using the HOPS task \texttt{fourfit}. Detections with high S/N were used to determine several important quantities for further processing. First, we derived the phase offsets between the 32 MHz channels within each band. Second, approximate atmospheric coherence times maximizing the S/N of detection were calculated to guide further incoherent fringe searching in the HOPS task \texttt{cofit}.  Third, the residual single-band delay, multi-band delay, and delay rate were used to set up narrow search windows for each source to assist in fringe finding.

A form of phase self-calibration was used to find fringes on baselines with low S/N, including long baselines (e.g., SP) and baselines including the single CARMA antenna.
The phased CARMA station is very sensitive and therefore can be used as a reference station to derive phase corrections to be applied to other antennas to remove rapid atmospheric phase fluctuations through baselines with the phased CARMA station. The fringe fitting was done on baselines to station F (i.e., FD, SF, and PF), and data were segmented at a $\sim$5 s cadence.
These phases were then removed from each station prior to coherent fringe fitting on the low-S/N baselines using \texttt{fourfit}, leading to much better coherence and detections with higher S/N.

Detected fringes were segmented at a cadence of 1 s and incoherently averaged to produce estimates of the correlation coefficients not biased due to the noise and the coherence loss \citep{Rogers1995}.  We confirmed that correlation coefficients derived with and without the phase-referencing technique were consistent, indicating that this phase-referencing technique does not bias our amplitude estimates.  In addition, segmented bispectra were also formed at a 10 s cadence and averaged to construct scan-averaged estimates of the closure phase.

The visibilities were calibrated as in \citet{Lu2013} \citep[see also][]{Fish2011,Lu2012}. Visibilities were a-priori calibrated by multiplying the VLBI correlation coefficient by the geometric mean of the System Equivalent Flux Density (SEFD) of the pair of antennas. 
Additional instrumental effects on the SMA were corrected \citep[see][for details]{Lu2013}.
Finally, visibilities were amplitude self-calibrated assuming that the intra-site VLBI baseline at CARMA (FD) measures the same total flux density as the CARMA interferometer.
In principle this assumption could be incorrect due to arcsecond-scale structure in the jet, which could produce the appearance of different correlated flux densities on different baselines within CARMA. However, M87 in 2012 March satisfies our assumption, as the arcsecond-scale jet was dominated by its unresolved (i.e., point-like) radio core, while the radio flux from extended components were $<1$ \% of the core flux.
Thus, the VLBI amplitudes measured on the intra-site FD baseline should be consistent with the core flux density measured with CARMA as a connected array.  For each scan, band and site, gains were calculated for each station to maximize self-consistency of the visibilities, including consistency of the calibrated FD flux density with the total flux density measured by CARMA.  Calibration errors of $5$\% have been added in quadrature to the random errors associated with the fringe search and estimation of the correlation coefficient on each baseline following previous observations \citep[e.g.,][]{Fish2011,Lu2012}. Note that we flagged data on scans when CARMA has a low phasing efficiency due to bad weather conditions, showing systematic losses in gain-calibrated amplitudes.

\section{Results}
\subsection{First detections of closure phases of M87}

\begin{figure}[t]
\centering{}\includegraphics[width=1\columnwidth]{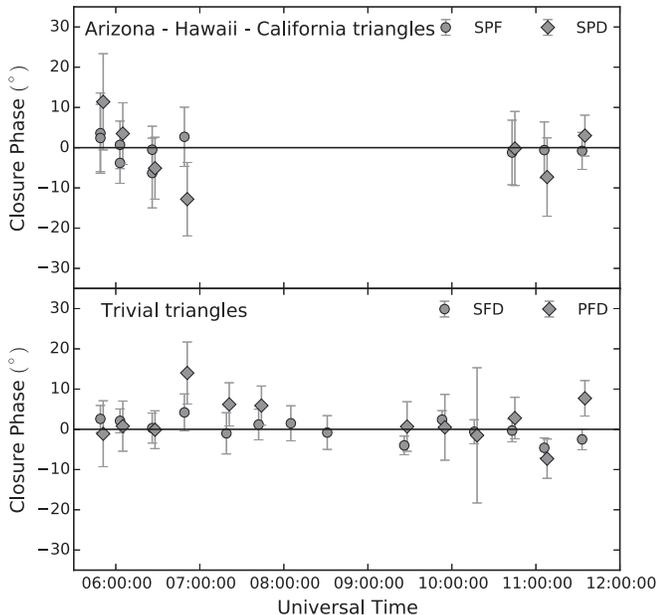}
\caption{
The measured closure phase of M87 as a function of time. Errors are 1$\sigma$. We added small offsets in UTC to each baseline data for avoiding overlaps of error bars.
Upper panel: the closure phase on AZ-CA-HI triangles.
Lower panel: the closure phase on trivial triangles, which include an intra-site baseline in CARMA. The closure phase on the trivial triangle is expected to be zero.
\label{fig:cphase}}
\end{figure}

\begin{table*}[t]
	\caption{Closure Phase of M87\label{tab:cphase}}
	\begin{center}
	\scriptsize
	\begin{tabular}{ccccccccccccc}
		\hline 
		\hline Year & DOY & \multicolumn{2}{l}{UTC} & Triangle & $u_{\rm XY}$ & $v_{\rm XY}$ & $u_{\rm YZ}$ & $v_{\rm YZ}$ & $u_{\rm ZX}$ & $v_{\rm ZX}$ &Closure & 1$\sigma$\\
		& & & & & & & & & & & Phase & error\\ 
		&  & (h) & (m) & (XYZ) & (M$\lambda$) & (M$\lambda$) & (M$\lambda$) & (M$\lambda$) & (M$\lambda$) & (M$\lambda$) & ($^\circ$) & ($^\circ$)\\ 
		\hline
		2012 & 81 & 5 & 49 & SPF & -2336.473 & -426.660 & 1997.198 & 847.338 & 339.275 & -420.677 & 3.60  9.98\\
		2012 & 81 & 6 &  3 & SPF & -2483.493 & -458.173 & 2113.293 & 874.212 & 370.200 & -416.039 & -3.80 & 5.05\\
		2012 & 81 & 6 & 26 & SPF & -2704.560 & -513.928 & 2286.622 & 921.497 & 417.939 & -407.569 & -6.30 & 8.67\\
		2012 & 81 & 5 & 49 & SPD & -2336.473 & -426.660 & 1997.150 & 847.284 & 339.323 & -420.624 & 11.40 & 11.96\\
		2012 & 81 & 6 &  3 & SPD & -2483.493 & -458.173 & 2113.246 & 874.158 & 370.247 & -415.985 & 3.50 & 7.65\\
		\hline
	\end{tabular}
	\normalsize
	\end{center}%
	(This table is available in its entirety in a machine-readable form in the online journal. A portion is shown here for guidance regarding its form and content.)
\end{table*}

We detected fringes on baselines to all 3 sites, consistent with the results of \citet{Doeleman2012}.  Furthermore, we detected closure phases on the Arizona-Hawaii-California triangle. Figure \ref{fig:cphase} shows the measured closure phase on the SPF/SPD triangles (upper; hereafter VLBI triangles) listed in Table \ref{tab:cphase} and SFD/PFD triangles (lower; hereafter trivial triangles). The average error bar on closure phases is $10.3{}^{\circ}$ for VLBI triangles and $5.0{}^{\circ}$ for trivial triangles. The error-weighted average of the closure phases by the square of S/N is $-0.7{}^{\circ}\pm2.9^{\circ}$ for VLBI triangles
and $-0.1^{\circ}\pm0.6^{\circ}$ for trivial triangles.
%, where errors are 1$\sigma$.
The closure phase is consistent with zero on trivial triangles, as would be expected if the source is point-like on arcsecond scales.
%supports its robust detection, since point-like structure of
%M87 on arc-second scale requires the zero closure phase on trivial
%triangles.
All closure phases on VLBI triangles coincide with zero within 1$\sigma$ level except 1 data point, which is consistent with zero within 2$\sigma$ levels. We note that non-detections in VLBI triangles during 7:00-10:00 UTC are attributable to non-detections on the SP baseline (Figure \ref{fig:amp}).

%The closure phase becomes 0$^{\circ}$, if the structure has point-symmetry. Considering the sharp triangle of current VLBI triangles, our results require symmetric structure on $\sim40$ ${\rm \mu}$as scales roughly along or perpendicular to the long baseline vectors from the US mainland to Hawaii. 

\subsection{The geometrical model of M87\label{subsec:geometrical_models}}

\begin{figure*}[t]
\centering{}%
\begin{tabular}{cc}
\begin{minipage}[t]{0.5\textwidth}%
\begin{center}
\includegraphics[width=1\textwidth]{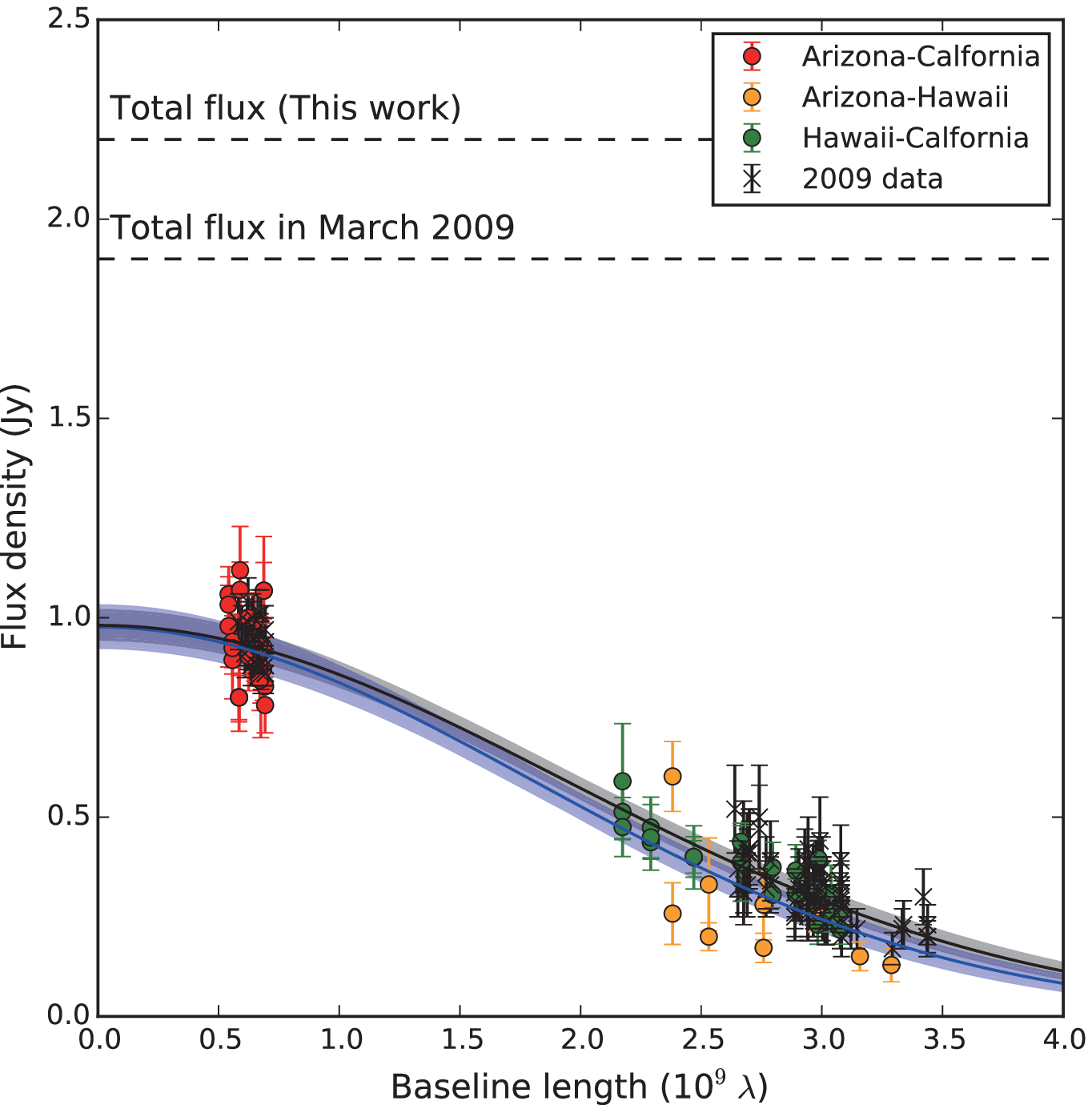}
\par\end{center}%
\end{minipage} & %
\begin{minipage}[t]{0.5\textwidth}%
\begin{center}
\includegraphics[width=1\textwidth]{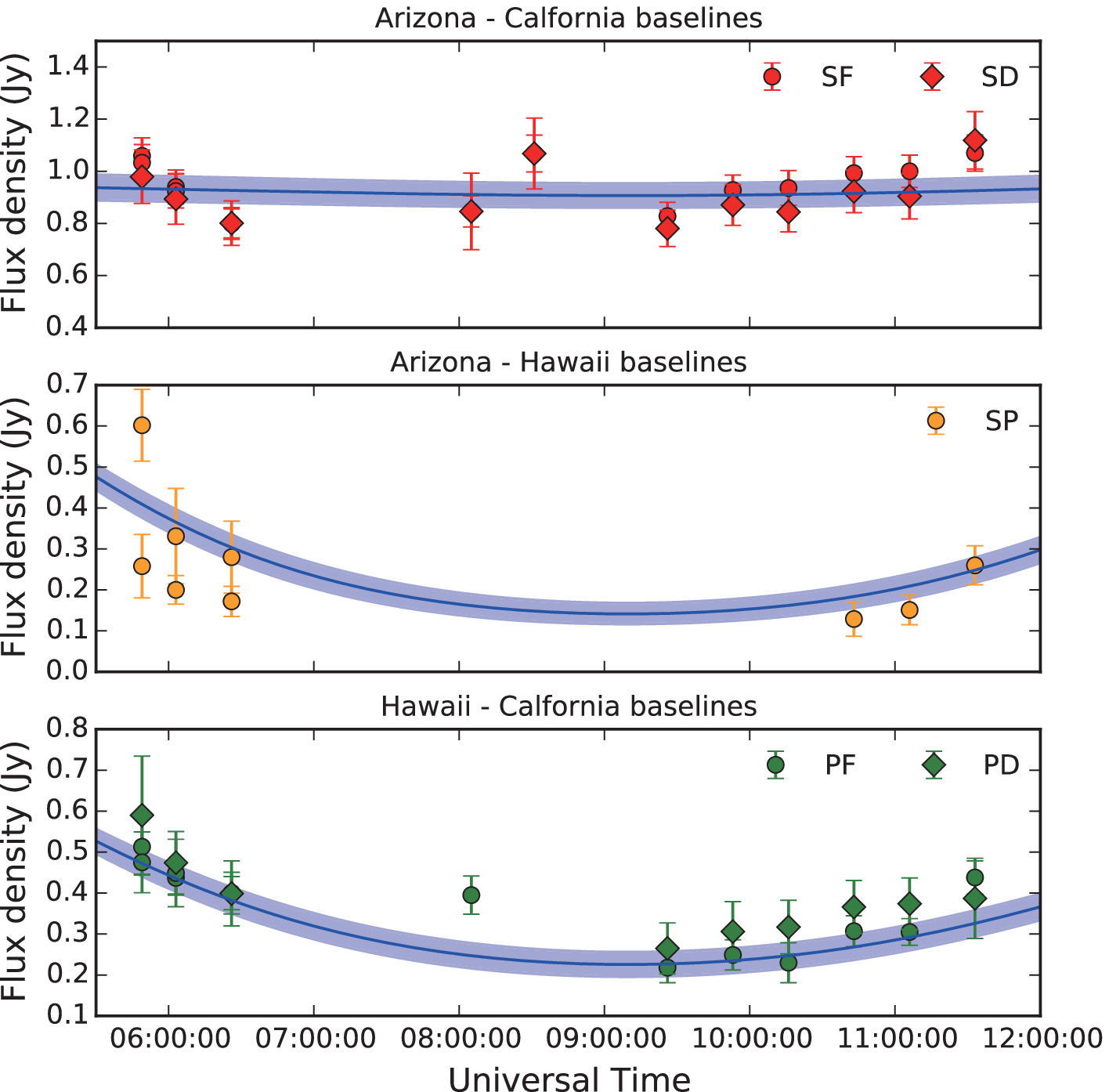}
\par\end{center}%
\end{minipage}\\
\end{tabular}
\caption{
The measured correlated flux density of M87. Circles and crosses indicate the correlated flux density observed in 2012 (This work) and 2009 \citep{Doeleman2012}, respectively. Errors are 1$\sigma$. The blue
line and light-blue region are best-fit models for the 2012 data and 3$\sigma$ uncertainties on it, respectively, while the black line and gray region are for the 2009 data.
Left panel: correlated flux density as a function of baseline length. 
Right panels: correlated flux density as a function of universal time for each baseline. 
\label{fig:amp}}
\end{figure*}

\begin{table*}[t]
	\caption{Gain-corrected Visibility Amplitudes of M87\label{tab:vis_amp}}
	\begin{center}
	\begin{tabular}{ccccccccc}
		\hline 
		\hline Year & DOY & \multicolumn{2}{l}{UTC} & Baseline & $u$ & $v$ & Correlated Flux & 1$\sigma$\\
		& & & & & & & Density & Error\\
		&     & (h) & (m) &          & (M$\lambda$) & (M$\lambda$) & (Jy) & (Jy)\\ 
		\hline
		2012 & 81 & 5 & 49 & PD &  2002.665 &  845.916 & 0.59 & 0.14\\
		2012 & 81 & 5 & 49 & PF &  2002.599 &  845.883 & 0.51 & 0.07\\
		2012 & 81 & 5 & 49 & PF &  2002.599 &  845.883 & 0.47 & 0.07\\
		2012 & 81 & 5 & 49 & SD &  -340.631 &  421.034 & 0.98 & 0.10\\
		2012 & 81 & 5 & 49 & SF &  -340.698 &  421.001 & 1.06 & 0.07\\
		2012 & 81 & 5 & 49 & SF &  -340.698 &  421.001 & 1.03 & 0.07\\
		2012 & 81 & 5 & 49 & SP & -2343.297 & -424.882 & 0.60 & 0.09\\
		2012 & 81 & 5 & 49 & SP & -2343.297 & -424.882 & 0.26 & 0.08\\
		\hline
	\end{tabular}
	\end{center}%
	(This table is available in its entirety in a machine-readable form in the online journal. A portion is shown here for guidance regarding its form and content.)
\end{table*}

The correlated flux density of M87 is shown in Figure \ref{fig:amp} and Table \ref{tab:vis_amp}. The arcsecond-scale core flux density of 2.2 Jy is $\sim$17 \% higher than the 1.9 Jy measured in 2009 (D12). This brightening on arcsecond scales is not accompanied by changes on VLBI scales. The visibility amplitudes are broadly consistent with 2009 results of D12, confirming the presence of the event-horizon-scale structure. This indicates that the region responsible for the higher flux density must be resolved out in these observations and therefore located somewhere down the jet.

The most of the missing flux on VLBI scales most likely attributes to the extended jet inside the arcsecond-scale radio core including the bright and stable knots such as HST-1. In the last decade, no radio enhancement was detected in such bright knots except the 2005 VHE flare at HST-1. Even for the exceptionally variable HST-1, the radio flux has been decreasing from the 2005 VHE flare to at least the end of the 2012 event (see \citealt{Abramowski2012} and H14). The observed increment in the missing flux seems incompatible with this trend in the bright knots, favoring that the missing flux originates in the vicinity of the radio core on milliarcsecond scales rather than the bright knot features. We discuss it in a physical context related with the 2012 event in \S\ref{subsec:implications_for_VHE_flare}.

%, with the exception that data obtained during 6:30-8:00 UTC appear to have a lower flux density on Arizona-Calfornia baselines. The phasing efficiency of CARMA was under $60$\% in this period owing to bad weather conditions, indicating that this lower flux density may be due to coherence losses.  Excluding these data, the measured correlated flux is broadly consistent with 2009 results.

The structure of M87 is not yet uniquely constrained, since millimeter VLBI detections of M87 remain limited in terms of baseline length and orientation, similar to previous observations in D12. Even with our detections of closure phase, our small data set is consistent with a variety of geometrical models (see \S\ref{subsec:physical_models} for physically motivated models). It is still instructive to investigate single-Gaussian models, which inherently predict a zero closure phase, to estimate the flux and approximate size of VLBI-scale structure and compare with the results of the previous observations.

\begin{table}[t]
\centering{}
\caption{Geometrical Models of M87. Errors are 3$\sigma$.\label{tab:models}}
\begin{tabular}{ccccc}
\hline 
\hline Model & Date & Compact Flux & FWHM & $\chi_{\nu}^{2}$ (d.o.f) \\
 &  & Density & Size & \\
 & (Year/DOY) & (Jy) & (${\rm \mu}$as) &\\
%\hline 
%\multicolumn{7}{c}{Circular Gaussian Model}\\
\hline 
2012 & 2012/81 & $0.98 \pm 0.05$ & $42.9 \pm 2.2$ & 2.2 (54)\\
2009\tablenotemark{a} & 2009/95-97 & $0.98 \pm 0.04$ & $40.0 \pm 1.8$ & 0.6 (102)\\
\hline 
\end{tabular}\tablenotetext{1}{Model obtained from all 3 days of data
  in the 2009 observations.}
\end{table}

Circular Gaussian fits to the visibility amplitudes on VLBI baselines are shown in Table \ref{tab:models}. The parameters of the best-fit circular Gaussian model agree with values obtained by D12. The compact flux density of
$0.98 \pm 0.05$ Jy is precisely consistent with the D12 value, while the size of $42.9 \pm 2.2$ $\mu$as (corresponding to $5.9 \pm 0.2$ $R_{{\rm s}}$) is slightly larger but still consistent within 3$\sigma$ uncertainty. We find no evidence of significant changes in event-horizon-scale structure between the 2009 and 2012 observations.

\section{Discussion}
\subsection{Physical models for the structure of 230 GHz emission\label{subsec:physical_models}}

Physically motivated structural models have been proposed for the Schwarzschild-radius-scale structure at 230 GHz in M87 for both jet and disk models \citep{Broderick2009,Dexter2012,Lu2014}. Although all proposed models predict the existence of a feature at the last photon orbit illuminated by a counter jet and/or accretion disk in the close vicinity of the black hole, there are significant differences between model images. The closure phase is an ideal tool to constrain physically motivated models, since relativistic effects such as gravitational lensing, light bending and Doppler beaming generally induce asymmetric emission structure at the vicinity of the central black hole, causing the closure phase to be nonzero.

\begin{figure*}[t]
\centering{}
\includegraphics[width=0.8\textwidth]{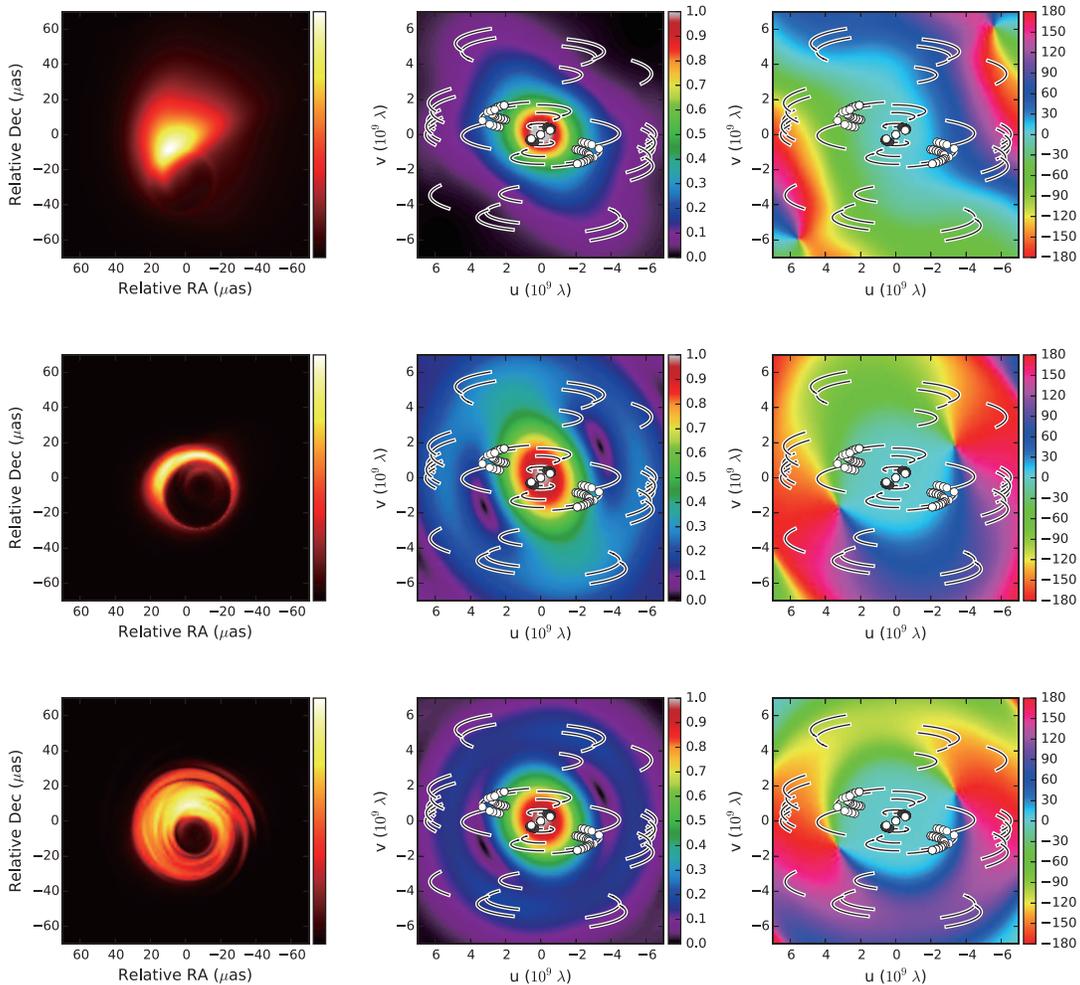}
\caption{
Images (left panels) and distributions of the visibility amplitude (middle panels) and visibility phase (right panels) of the physical models for the structure of 230 GHz emission. The white circle points shows the $uv$-coverage of our observations, while the black lines show the $uv$-coverage of future observations with current US stations, LMT in Mexico, IRAM 30m telescope in Spain, PdBI in France and ALMA/APEX in Chile. (Top panels) an approaching-jet-dominated model \citep{Broderick2009,Lu2014} fitted to 2009 data in \citet{Doeleman2012} (Broderick et al. in prep.). (Middle panels) a counter-jet-dominated model (J2) in \citet{Dexter2012} at a position angle of $-70^\circ$ inferred for the large-scale jet. (Bottom panels) an accretion-disk-dominated model (DJ1) in \citet{Dexter2012} at a position angle of $-70^\circ$ inferred for the large-scale jet.
\label{fig:images}}
\end{figure*}

%explanation of each image
Figure \ref{fig:images} shows images and visibilities of the approaching-jet-dominated models \citep{Broderick2009,Lu2014}, counter-jet-dominated models \citep[J2 in][]{Dexter2012}, and the accretion-disk-dominated models \citep[DJ1 in][]{Dexter2012}. For jet models, 230 GHz emission structure can be categorized into two types. One is the approaching-jet-dominated models, where emission from the approaching jet is predominant at 230 GHz \citep{Broderick2009,Lu2014}. The model images consist of bright blob-like emission from the approaching jet and a weaker crescent or ring-like feature around the last photon orbit illuminated by a counter jet. The emission from the approaching jet dominates the 230 GHz emission regardless of the loading radius of non-thermal particles where leptons are accelerated and the jet starts to be luminous, although the crescent-like feature appears more clearly at smaller particle loading radii \citep[see Figure 3 in][]{Lu2014}. In counter-jet-dominated models, the counter-jet emission is predominant instead of the approaching jet. Such a situation could happen if the bright emission region in the jet is very close to the central black hole (within few $R_{\rm s}$) suppressing the approaching jet emission due to gravitational lensing. Photons from the counter jet illuminate the last photon orbit, forming a crescent-like feature. It is worth noting that \citet{Dexter2012} and \citet{Lu2014} have clear differences in their images even at the same particle loading radius of a few  $R_{\rm s}$, most likely due to differences in magnetic-field distribution and also the spatial and energy distribution of non-thermal particles in their models.  The accretion disk models are well characterized by a crescent-like or ring-like feature around the last photon orbit. The 230 GHz emission arises in the inner portion of the accretion flow ($r\sim 2.5$ $R_{\rm s}$) near the mid-plane.
%In both the counter-jet-dominated and accretion-disk-dominated models, the brightest region is offset from the event horizon in direction perpendicular to the jet axis.

%\begin{figure*}[t]
%\centering{}%
%\begin{tabular}{cc}
%\begin{minipage}[t]{1\columnwidth}%
%\begin{center}
%\includegraphics[width=1\textwidth]{figure5a.eps}
%\par\end{center}%
%\end{minipage} & %
%\begin{minipage}[t]{1\columnwidth}%
%\begin{center}
%\includegraphics[width=1\textwidth]{figure5b.eps}
%\par\end{center}%
%\end{minipage}\\
%\end{tabular}
%\caption{
%The closure phases of models J2 (left) and DJ1 (right) for various jet position angles ranging from -110$^\circ$ to -40 $^\circ$ in  \citet{Dexter2012} as a function of Greenwich Sidereal Time.
%Upper panel: model closure phases on the current VLBI triangle. The circular points are our results shown in Figure \ref{fig:cphase}.
%Middle panel: model closure phases on a triangle including SMA in Hawaii, CARMA in California and LMT in Mexico.
%Lower panel: model closure phases on a triangle including SMA in Hawaii, CARMA in California and ALMA in Chile.
%\label{fig:cphase_jason}}
%\end{figure*}

\begin{figure}[t]
\centering{}\includegraphics[width=0.8\columnwidth]{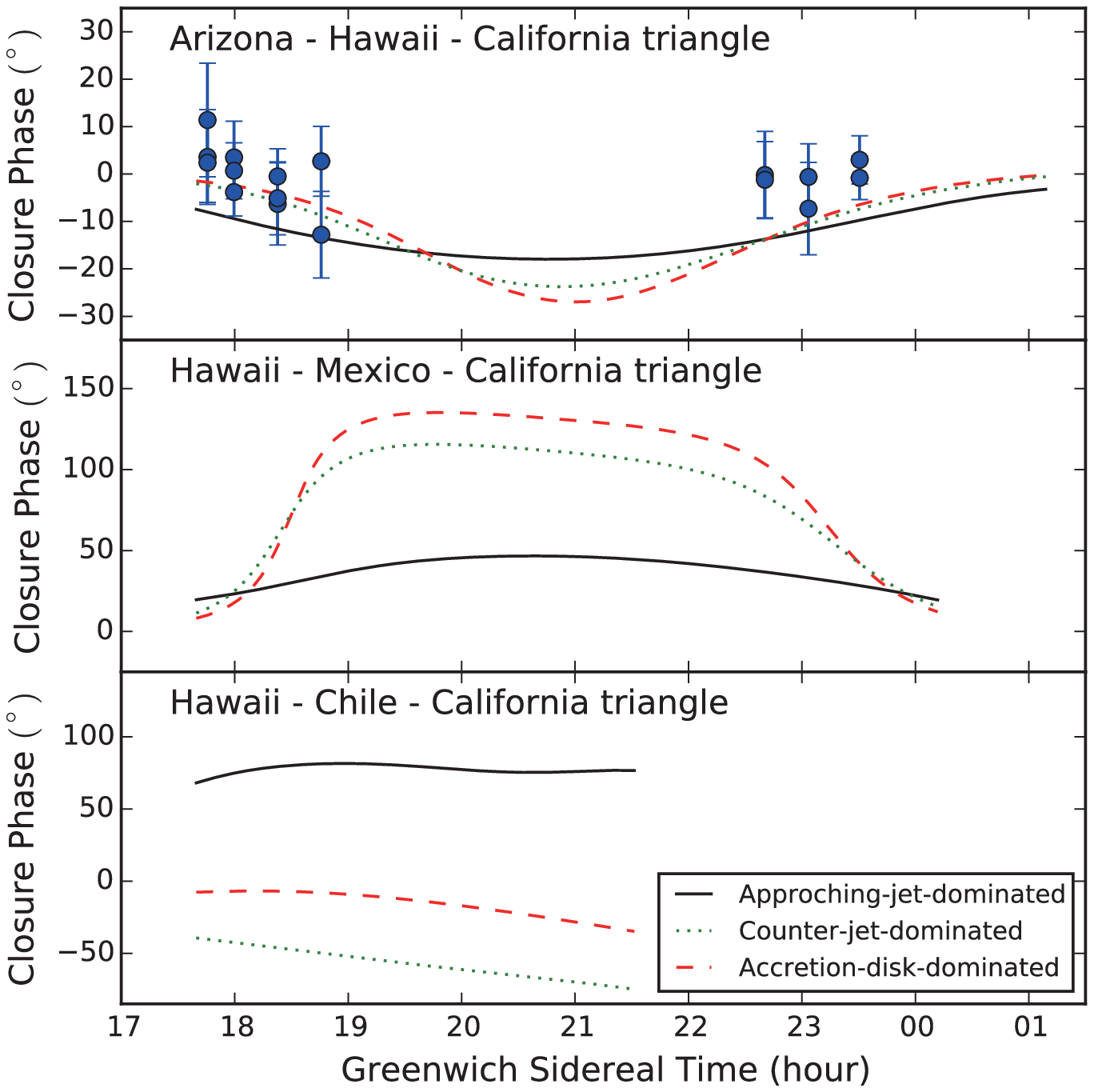}
\caption{
The closure phase of models in Figure \ref{fig:images} as a function of Greenwich Sidereal Time. The black solid-line shows an approaching-jet-dominated model \citep{Doeleman2008,Lu2014} fitted to 2009 data in \citet{Doeleman2012} (Broderick et al. in prep.). The red dashed- and green dotted-lines indicate accretion-disk-dominated and counter-jet-dominated models (DJ1 and J2) in \citet{Dexter2012}, respectively, at a position angle of $-70 ^\circ$ inferred for the large-scale jet.
Upper panel: model closure phases on the current VLBI triangle. The circular points are our results shown in Figure \ref{fig:cphase}.
Middle panel: model closure phases on a triangle including SMA in Hawaii, CARMA in California and LMT in Mexico.
Lower panel: model closure phases on a triangle including SMA in Hawaii, CARMA in California and ALMA in Chile. (A color version of this figure is available in the online journal.)
\label{fig:cphase_avery}}
\end{figure}

% quantitative comparison
Measured closure phases on the Hawaii-Arizona-California triangle are consistent with these three models. In Figure \ref{fig:cphase_avery}, we show the model closure phases calculated in the MIT Array Performance Simulator\footnote{http://www.haystack.mit.edu/ast/arrays/maps/} (MAPS) for an approaching-jet-dominated model, a counter-jet-dominated model and an accretion-disk-dominated model in Figure \ref{fig:images}.
The closure phase of the approaching-jet-dominated model is almost zero. On the other hand, the model closure phases of counter-jet-dominated and accretion-disk-dominated models are systematically smaller than observed closure phase in the later GST range, but the models and observed closure phases are consistent within a 3-sigma level. We note that the results for counter-jet-dominated and accretion-disk-dominated models shown in Figure 4 disagree with Figure 9 of \citet{Dexter2012}, due to a mistake in \citet{Dexter2012} in constructing the closure phase triangles.
%\textcolor{red}{We also note that} the closure phase \textcolor{red}{of DJ1 and J2} is still consistent at various position angles (PAs) ranging from -110$^\circ$ to -40$^\circ$ inferred for the position angles at the \textcolor{red}{43 GHz} radio core \citep{Hada2011}.

% closure phase
All three models commonly predict small closure phases on the Hawaii-Arizona-California triangle. Visibility phases on the Arizona-California baseline, which barely resolves the source, are nearly zero.  The closure phase on current VLBI triangles are almost same to differences in the visibility phase between long baselines between Hawaii and the US mainland. For the case of the approaching-jet-dominated models, the phase gradient between long baselines is expected to be moderate particularly at large particle loading radii, since emission is blob-like and fairly symmetric on spatial scales corresponding to the current long baselines. Models with a clear crescent-like or ring-like feature generally predict a steep phase gradient around the null amplitude region (see Figure \ref{fig:images}), which would be detectable not on the current baselines but more longer baselines such as the Hawaii-Mexico baseline.
%, since the 230 GHz structure is hardly resolved and then the visibility phase is very close to zero on these baselines as shown in Figure \ref{fig:images}. Newly detected closure phases suggest that the visibility phase on long baselines is nearly canceled out, requiring a flat phase gradient around spatial frequencies sampled by long baselines between the US mainland and Hawaii.

% qualitative explanation for the theoretical prediction
%More generally, these classes of physical models commonly predict closure phases close to zero on current VLBI triangles. 
%which can cause a large phase difference even between two close long baselines between the US mainland and Hawaii. However, the baseline length of current long baselines is still shorter than the first null region. We note that all above models adopt the higher black-hole mass of $\sim 6\sim 10^9$ M$_{\rm solar}$, but such an effect of the ring-like or crescent-like feature is not significant also in the lower mass case, since the lower black-hole mass produces more compact shadowed region in the image plane and the first null region at a longer baseline length in the visibility plane. Thus, these physical models predict that current VLBI triangles are too short to detect significant asymmetry in the structure regardless of the existence of the bright last photon orbit, and then expect the closure phase close to zero on those triangles.

The observed closure phases cannot distinguish between models with different dominant origin of 230 GHz emission on the current VLBI triangle due to the large errors on the data points. However, future observations with a higher recording rate of 16 Gbps can measure the closure phase with an accuracy within a few degrees at 1 minute integration, which can constrain physical models more precisely. In addition, models can be effectively distinguished by near future observations with additional telescopes such as the Large Millimeter Telescope (LMT) in Mexico or the Atacama Large Millimeter/submillimeter Array (ALMA) in Chile.  Differences in closure phases between models become more significant on larger triangles, as shown in the middle and bottom panels of Figure \ref{fig:cphase_avery}. 

While all of the physically motivated models are broadly consistent with the currently measured closure phases and amplitudes, they make dramatically different predictions for forthcoming measurements. Models in which the image is dominated by contributions close to the horizon (counter-jet-dominated and accretion-disk-dominated models) exhibit large closure phases on triangles that include the LMT in stark contrast to those dominated by emission further away (approaching-jet-dominated). This extends to the visibility amplitudes: the compact emission models predict nulls on baselines probed by ALMA and the LMT (see Figure \ref{fig:images}).

\subsection{The brightness temperature of the event-horizon-scale structure \label{subsec:brightness temperture}}
New VLBI observations of M87 at 230 GHz in 2012 confirm the presence of the event-horizon-scale structure reported in D12. The compact flux density and effective size derived from the circular Gaussian models allow us to estimate the effective brightness temperature of this structure, which is given by \citep[e.g.][]{Akiyama2013}
\begin{eqnarray}
T_b &=& \frac{c^2}{2k_B \nu^2} \frac{F}{\pi \phi ^2 /4\ln 2}\\
    &=& 1.44 \times 10^{10} \,{\rm K}\times \left( \frac{\nu}{\rm 230 \,GHz} \right)^{-2}
    \left( \frac{F}{\rm 1\, Jy}\right)\left( \frac{\phi}{\rm 40 \, \mu as}\right)^{-2},
\end{eqnarray}
where $F$, $\nu$ and $\phi$ are the total flux density, observation frequency, and the FWHM size of the circular Gaussian. The effective brightness temperature is $1.42^{+0.11}_{-0.10} \times 10^{10}$ K for the 2009 model and $(1.23 \pm 0.11) \times 10^{10}$ K for the 2012 model, where errors are 3 $\sigma$. These brightness temperatures of $\sim 2\times 10^{10}$ K are below the upper cutoff in the intrinsic brightness temperature of $\sim 10^{11}$ K on the "inverse Compton catastrophe" argument (e.g. \citealt{Kellermann1969,Blandford1979,Readhead1994}). Although the 1.3 mm VLBI structure has been poorly constrained particularly N-S direction possibly inducing additional uncertainties, it is still instructive to discuss the effective brightness temperature and its physical implications for both the jet and accretion disk scenario.

In the case of the jet scenarios, the brightness temperature would not be highly affected by the Doppler beaming, and then not significantly differ from the intrinsic (i.e. not Doppler-boosted) brightness temperature. The brightness temperature is amplified by a factor of $\delta$ for an isotropic blob-like source \citep[e.g.][]{Urry1995}. The Doppler factor is $1 - 3$ at a moderate viewing angle of $15 - 25^\circ$ \citep[e.g.][]{Hada2011} and the Lorentz factor of $1-2$ in the inner 10$^2$ $R_{\rm s}$ region \citep[e.g.][]{Asada2014} inferred for the M87 jet.

%The brightness temperature is a good indicator of the Doppler factor of the jet. The brightness temperature of $\sim 10^{10}$ K does not exceed the upper cutoff in the intrinsic (i.e. not Doppler-boosted) brightness temperature of $\sim 10^{11}$ K on the "inverse Compton catastrophe" argument (e.g. \citealt{Kellermann1969,Blandford1979,Readhead1994}). The observed brightness temperature of the radio core exceeds this upper cutoff in many sources with a high bulk Lorentz factor of $\gtrsim 10$ \citep[e.g.][]{Horiuchi2004}, since the brightness temperature is amplified by a factor of $\delta$. The low brightness temperature of $\sim 10^{10}$ K on 230 GHz emission in M87 \textcolor{red}{seems incompatible with theoretical models expecting the high Lorentz factor}

Interestingly, the 230 GHz brightness temperature is broadly consistent with the peak brightness temperature of $\sim 10^{9} - 10^{10}$ K at the radio cores at lower frequencies from 1.6 GHz to 86 GHz \citep[e.g.][]{Dodson2006,Ly2007,Asada2012,Hada2012,Nakamura2013} located within $ \sim 10^2 $ $R_{\rm s}$ from the jet base \citep{Hada2011}. 
This would provide some implications also for the magnetic field structure of the jet. If we assume the radio core surface corresponds to the spherical photosphere of the synchrotron self-absorption at each frequency, the magnetic field strength at the radio core can be estimated by \citep[e.g.][]{Marscher1983,Hirotani2005,Kino2014}
\begin{eqnarray}
B &=& b(p)\nu^5\phi ^4F ^{-2}\frac{\delta}{1+z} \propto \nu T_b ^{-2} \frac{\delta}{1+z}.
\end{eqnarray}
The constant Doppler factor and brightness temperature give the magnetic field strength roughly proportional to the observation frequency at the radio core (i.e. $ B_{\rm core} \propto \nu _{\rm obs}$). Using the frequency-dependence of the radio core position \citep[$r _{\rm core} \propto \nu_{\rm obs}^{-0.94 \pm 0.09}$; ][]{Hada2011}, the magnetic field strength at the radio core is inversely proportional to the distance from the jet base approximately (i.e. $ B_{\rm core} \propto r _{\rm core}^{-1}$) in inner $\sim 10^2$ $R_{{\rm s}}$. 
This magnetic field profile can be obtained if the transverse (i.e. nearly toroidal) magnetic field dominates on this scale rather than the longitudinal (i.e. nearly poloidal) field along the conical stream with no velocity gradient under the flux frozen-in condition (\citealt{Blandford1979}; also see \S5 in \citealt[][]{Baum1997}). This profile also can be obtained if the longitudinal (i.e. nearly poloidal) field dominates the transverse (i.e. nearly toroidal) magnetic field along the paraboloidal stream under the flux frozen-in condition, although recent observations favor a conical stream of the jet in inner $\sim 10^2$ $R_{{\rm s}}$ \citep{Hada2013}.

Even though above assumptions might not work well for M87, this simple analysis suggests that the dominance of toroidal or poloidal magnetic fields starts to become a major concern on the jet formation in inner $\sim 10^2$ $R_{{\rm s}}$. Future EHT observations with additional stations and space VLBI observations \citep[e.g. Radio Astron;][]{Kardashev2013} will provide more detailed structure of the radio core including the profile of the stream line, enabling more precise analysis on the magnetic field structure of the relativistic jet in M87.

The measurements of the brightness temperature also give some implications for the energetics of the jet base. The equipartition brightness temperature \citep{Readhead1994} of the non-thermal plasma with the flux density of $ \sim 1 $Jy at 230 GHz is $T_{\rm eq} \leq 10^{12} $ K, where the equality is given if 230 GHz emission is fully optically thick. This gives the ratio between the internal energy of non-thermal leptons and the magnetic-field energy density $U_B/U_e=T_{\rm eq}/T_b \leq 10^2$ \citep{Readhead1994}. This implies that, if the 230 GHz emission is dominated by optically-thick non-thermal synchrotron emission, the magnetic-field energy dominates the internal energy of the non-thermal particles at the jet base. We note that, recently, \citet{Kino2015} performed more careful analysis on the energetics at the jet base, stating that the magnetic-field energy is dominant even in fully optically-thin case unless protons are relativistic.

The brightness temperature is broadly consistent with the electron temperature of $\sim 10^{9-10}$ K as expected for RIAF-type accretion disks \citep[e.g.][]{Manmoto1997,Narayan1998,Manmoto2000,Yuan2003}. The brightness temperature is a factor of $\sim 2-3$ smaller than that of Sgr A* with similar size and higher flux density \citep[][]{Doeleman2008,Fish2011}. If the 230 GHz emission is dominated by thermal synchrotron emission from the accretion disk in both Sgr A* and M87, it seems broadly consistent with a theoretical prediction that a disk with higher accretion rate has a lower electron temperature due to enhanced electron cooling \citep[see Fig.2 in][]{Mahadevan1997}.

\subsection{Implications for the VHE enhancement in March 2012 \label{subsec:implications_for_VHE_flare}}
%Numerous physical models have been proposed to explain observed activity in VHE emission \citep[see][for a review]{Abramowski2012}. Thus, we briefly discuss the general implications of our results and those of H14 on the 2012 VHE event.

%They generally predict a compact region for VHE emission ascribable to rapid VHE activities on timescales of $\lesssim$ 1 day.  However, the 2012 event exhibited activity on much longer timescales (see BV12).They generally predict a compact region for VHE emission ascribable to rapid VHE activities on timescales of $\lesssim$ 1 day. 

Our observations were performed in the middle of the VHE enhancement reported in \citet{Beilicke2012} and H14. There are several observational pieces of evidence for the existence of a radio counterpart to the VHE enhancement around our observations. First, the onset of the radio brightening at 22 and 43 GHz occurs simultaneously with the VHE enhancement, indicating that the radio and VHE emission regions are not spatially separated. Since the radio brightening starts $\sim$ 20-30 days before our observations, 230 GHz emission is also expected to be enhanced at the epoch of our observations. The radio flux measured with the SMA in H14 indeed shows a local maximum in its light curve during our observations, which is consistent with our results showing a radio flux greater than in April 2009 on arcsecond scales when M87 was in a quiescent state \citep{Abramowski2012}. Second, the radio counterpart was not resolved in the radio core in VERA observations, suggesting that the radio counterpart of the 2012 event should exist near the radio core at 43 GHz located at a few tens of Schwarzschild radii downstream from the central black hole and/or the jet base visible at 230 GHz.

The geometrical model (described in \S\ref{subsec:geometrical_models}) suggests that there are no obvious structural changes on event-horizon scales between 2009 and 2012, despite the increase in the core flux on arcsecond scales. One plausible scenario for explaining the different behavior between event-horizon scales and arcsecond scales is that the structure of the flare component at 230 GHz has extended structure that is resolved out with the current array. The shortest VLBI baseline in our observations, SMT-CARMA, has a length of 600 M$\lambda$. If we consider the Gaussian-like structure for the flaring region with a radio flux of few $\times$ 100 mJy corresponding to the flux increment at the local peak in the 230 GHz light curve of H14, the flaring region should be extended enough to have a correlated flux smaller than the standard deviation on SMT-CARMA baselines of $\sim 90$ mJy so that the increment in the radio flux is not significantly detected on those baselines. The minimum FWHM size can be estimated to be $\sim 140$ $\mu$as $\sim20$ $R_{\rm s}$, which has a HWHM size of $ \sim 600  $  M$\lambda$ in the visibility plane. 
%, corresponding to an angular scale of $\sim$ 0.3 mas $\sim$ 41 $R_{{\rm s}}$.
This limitation is consistent with at least two aspects of VHE flares.

%First, this scenario agrees with detected closure phases consistent with zero, indicating that there is no additional component with different size and flux density. The visibility amplitudes could be explained by the sum of the ISCO scale structure detected in D12 with a lower flux density and an additional compact flaring component. However, such a structure would be inconsistent with the closure phase results if they are spatially separated.  Although zero closure phase could be produced if they are spatially unresolved enough to be approximately concentric, it requires at least an additional explanation why the flaring component stays close to the centroid of the event-horizon-scale structure with a radius of $\sim 2$ light days even $\sim$ 20-30 days after the start of the enhancement at both \textcolor{red}{22 and 43 GHz} and the VHE band, implying a very slow bulk speed of $\ll 0.1c$. In addition, it would be incompatible with the frequency-dependent position of the radio core \citep{Hada2011}, requiring that the flare component reach a few tens of Schwarzschild radii downward from the jet base for the flare to be detected at \textcolor{red}{22 and 43 GHz}.

First, the 2-month duration of the 2012 VHE event implies that the size of the emission region is $<60\delta$ light days $\sim 0.6\delta$ mas, from causality considerations. Similar constraints of $<0.44$ mas $\sim 60 R_{{\rm s}}$ are provided with VERA at 43 GHz in H14, since the flare component was not spatially resolved during their observations. Combining our measurement with these size limits, the VHE emission region size during our observations is constrained to be in the tight range of $\sim 0.14-0.44$ mas, corresponding to $\sim$20 - 60 $R_{{\rm s}}$.

Second, when the emission region size is larger than $\sim$ 20 $R_{{\rm s}}$, the emitted VHE photons will not be affected by absorption due to the process of photon-photon pair creation (${\rm \gamma\gamma}$-absorption). In principle, $\rm \gamma$-ray photons with energy $E$ interact most effectively with target photons in the infrared (IR) and optical photon field of energy \citep[e.g.][]{Rieger2011}
\begin{eqnarray}
\varepsilon (E) \sim \left( \frac{E}{1\,{\rm TeV}} \right) ^{-1}\,{\rm eV}.
\end{eqnarray}
Since the 2012 enhancement was detected at $\sim 0.3-5$ TeV in VHE regime \citep[see][]{Beilicke2012}, the target photon wavelength is $\sim 0.4-6$ $\rm \mu$m in the near-infrared (NIR) and optical regimes.
The optical depth of $\rm \gamma$-rays of energy $E$ for the center of an infrared source with a size $R$ and luminosity $L(\varepsilon )$ can be written by \citep[e.g.][]{Neronov2007}
\begin{eqnarray}
\tau _{\rm \gamma \gamma}(E,R) & \eqsim & \frac{\sigma_{T}}{5}\frac{L(\varepsilon (E) )}{4\pi R^{2}c\varepsilon}R \nonumber\\
& \eqsim & 0.25
\left( \frac{L \{ [E/(1\,{\rm TeV})  ] \, {\rm eV} \}}{10^{40}\,{\rm erg\,s^{-1}}} \right)\nonumber\\
& &\times \left( \frac{R}{20\,R_{s}} \right)^{-1}
\left( \frac{E}{5\,{\rm TeV}} \right). \label{eq:gamma-gamma}
\end{eqnarray}
The NIR and optical luminosity is $L\sim 10^{40}$ erg s$^{-1}$ within a few tens of parsecs at the nucleus \citep[e.g.][]{Biretta1991,Boksenberg1992}. Even in the extreme case that the flaring region accounts for all nucleus emission in the NIR and optical regime, the optical depth is smaller than unity at $E<$ a few TeV, where the enhancement was detected in \citet{Beilicke2012}, for the size of $\sim 20$ $R_{{\rm s}}$. This allows ${\rm \gamma}$-ray photons up to a few TeV to escape from the vicinity of the black hole, explaining why the 2012 event was detectable without introducing a special geometry of emission regions. Note that more careful calculation increases $\tau _{\rm \gamma \gamma}$ by a factor of several (\citealt{Brodatzki2011}; Broderick \& Tchekhovskoy in prep.), but even in this case the optical depth is smaller than unity for the upper half of the size range ($\sim 40-60$ $R_{{\rm s}}$).
%MEMO MIR observations: Perlman2001,Whysong2004,Reunanen2010,Asmus2011

The scenario limiting the size to a range of $\sim$20-60 $R_{{\rm s}}$ during our observations in the middle of the 2012 event can naturally explain our results and other observational results. It is instructive to compare this scenario to the numerous physical models proposed for the VHE emission in M87 \citep[see H14 and][for a review]{Abramowski2012}. Here, we briefly discuss general implications for the existing VHE models of M87 based on our scenario.

The size of $\sim$20-60 $R_{{\rm s}}$ is presumably incompatible with many existing models assuming extremely compact regions of $\lesssim$ 1-10 $R_{\rm s}$, ascribing the VHE emission to particle acceleration in the BH magnetosphere \citep[e.g.][]{Neronov2007,Rieger2008,Levinson2011,Vincent2014}, multiple leptonic blobs in the jet launch/formation region \citep[e.g.][]{Lenain2008}, leptonic models involving a stratified velocity field in the transverse direction \citep{Tavecchio2008}, mini-jets within the main jet \citep[e.g.][]{Giannios2010,Cui2012}, and interactions between a red giant star/gas cloud and the jet base \citep[e.g.][]{Barkov2010,Barkov2012}.
These models can reasonably explain the very short variable time scale of $\lesssim$ 1 d in the past three flares in 2005, 2008 and 2010, but are not favored for this particular event in 2012.
%Note that these models can reasonably explain the very short variable time scale of $\lesssim$ 1 d in the past three flares in 2005, 2008 and 2010, but it is not clear whether they can explain the long variability time scale of the 2012 event.

Consistency with the size limitation is less clear for models assuming different emission regions or different kinds of emitting particles for radio and VHE emissions, such as hadronic models \citep[e.g.][]{Reimer2004,Barkov2010,Reynoso2011,Barkov2012,Cui2012,Sahu2013} and some multi-zone leptonic models with a stratified velocity field in radial or transverse directions of the jet by involving the deceleration flow or the spine-layer structure, respectively \citep{Georganopoulos2005,Tavecchio2008}. Since the relation between radio and VHE emission has not been well formulated for these models, more detailed predictions particularly on the radio-TeV connection are required for further discussions.

Interestingly, a homogeneous one-zone synchrotron self-Compton (SSC) model (i.e. the standard leptonic model) predicts a comparable source size ($\sim$ 0.1 mas) to our scenario for a broadband SED in a relatively moderate state \citep{Abdo2009}. It also can naturally explain the radio-VHE connection in H14. The simple leptonic one-zone SSC model seems more plausible than other existing models for M87 to explain some properties such as the size constraint and the radio-VHE connection, but further dedicated modeling for the 2012 event would be required to test consistency with overall observational properties such as the broadband SED, which is not discussed here. Note that leptonic models might be problematic for explaining the hard VHE spectrum, which is common in the previous three VHE flares and the 2012 events, against the Klein-Nishina and $\rm \gamma$-ray opacity effects softening the VHE spectrum \citep[see, discussions in][]{Tavecchio2008}. 

Our new observations clearly show that short-mm VLBI is an useful tool to constrain the size of the radio counterpart, which is a new clue to understand the VHE activities in M87. In particular, new constraints can be obtained by combining simultaneous EHT observations with measurements of VHE spectra at $\gtrsim$ 10 TeV highly affected by ${\rm \gamma\gamma}$-absorption (see, Eq.(\ref{eq:gamma-gamma})) with the Cherenkov Telescope Array \citep[CTA,][]{Actis2011}.
In addition, the complementary dedicated monitoring with lower frequency monitoring on mas and arcsec scales is also important to study details on radio/VHE connections and also constrain on the important physical parameters.

\section{Summary}
New VLBI observations of M87 at 230 GHz in 2012 confirm the presence of the event-horizon-scale structure reported in D12. We summarize our results as follows;
\begin{enumerate}
\item For the first time, we have acquired 230 GHz VLBI interferometric phase information on M87 through measurement of closure phase on the triangle of long baselines. Measured closure phases are consistent with 0$^{\circ}$, as expected by physically-motivated models for 230 GHz structure such as jet models and accretion disk models. Although our observations can not currently distinguish models, we show that the future closure phase/amplitude measurements with additional stations and greater sensitivity can effectively distinguish and put a tight constrain on physical models.

\item The brightness temperature of the event-horizon-scale structure is $\sim 1 \times 10^{10}$ K both for previous observations (D12) and our new observations. This brightness temperature is broadly consistent with that of the radio core at lower frequencies from 1.6 to 86 GHz located in the inner $\sim 10^2$ $R_{{\rm s}}$. We demonstrated a simple analysis assuming that the observed radio core is the photosphere of synchrotron self-absorption. It shows that the constant brightness temperature may give the magnetic-field profile of $B\propto r_{\rm core}^{\sim -1}$ in inner $\sim 10^2$ $R_{{\rm s}}$, consistent with a prediction of the conical jet with no velocity gradient dominated by the toroidal magnetic field. This indicates that more precise imaging of the radio core with future EHT and space VLBI can address the magnetic field profile in inner $\sim 10^2$ $R_{{\rm s}}$ crucial for understanding the jet formation.

%\item Our results reveal that the VLBI structure of M87 in 2012 March does not change from the previous 2009 observations (D12), indicating that the structure on ISCO scales could be stable, contrary to some theoretical predictions of structural variations on year time-scales. We need to continue observations of M87 with the Event Horizon Telescope to test the existence of the structural variability on the event-horizon-scale structure.

\item Our observations were conducted in the middle of a VHE enhancement originating in the vicinity of the central black hole. The effective size derived from our data and results of lower-frequency observations favor the relatively extended size of VHE emission region of $\sim$20-60 $R_{{\rm s}}$. This would not favor VHE emission models that predict a compact emission region of $\lesssim$ 10 $R_{{\rm s}}$ for this event.
\end{enumerate}
It is clear that future VLBI observations with better sensitivity and additional baseline coverage will be crucial to constrain models of M87 on event-horizon scales.

\acknowledgments 
We appreciate an anonymous referee for helpful comments and constructive suggestions. 
K. Akiyama thanks Dr. Akihiro Doi, Prof. Alan Marscher, Dr. Svetlana Jorstad and Dr. Jose L. G\'{o}mez for fruitful discussions on scientific interpretations. 
K. Akiyama and K.H. are supported by a Grant-in-Aid for Research Fellows of the Japan Society for the Promotion of Science (JSPS). 
A.E.B. receives financial support from the Perimeter Institute for Theoretical Physics and the Natural Sciences and Engineering Research Council of Canada through a Discovery Grant.
L.L and G.O-L acknowledge the support of DGAPA, UNAM, and of CONACyT (M\'{e}xico).

Event Horizon Telescope work at the MIT Haystack Observatory and the Harvard Smithsonian Center for Astrophysics is supported by grants from the National Science Foundation (NSF) and through an award from the Gordon and Betty Moore Foundation (GMBF-3561).
The Arizona Radio Observatory (ARO) is partially supported through the NSF University Radio Observatories (URO) program under grant No. AST 1140030.
The Submillimeter Array is a joint project between the Smithsonian Astrophysical Observatory and the Academia Sinica Institute of Astronomy and Astrophysics and is funded by the Smithsonian Institution and the Academia Sinica.
Funding for ongoing CARMA development and operations is supported by the NSF and the CARMA partner universities.
Event Horizon Telescope work at the Mizusawa VLBI Observatory is financially supported by the MEXT/JSPS KAKENHI Grant Numbers 24540242, 25120007 and 25120008. Research at Perimeter Institute is supported by the Government of Canada through Industry Canada and by the Province of Ontario through the Ministry of Research and Innovation. This work has benefited from open source technology shared by the Collaboration for Astronomy Signal Processing and Electronics Research (CASPER).

{\it Facilities:} \facility{CARMA}, \facility{JCMT}, \facility{SMA} \& \facility{SMT}

\end{document}